\documentstyle[pre,preprint,aps,epsfig]{revtex}

\tightenlines

\newcommand{\bq}{\begin{equation}}
\newcommand{\eq}{\end{equation}}
\newcommand{\bqa}{\begin{eqnarray}}
\newcommand{\eqa}{\end{eqnarray}}
\newcommand{\ben}{\begin{enumerate}}
\newcommand{\een}{\end{enumerate}}
\newcommand{\bc}{\begin{center}}
\newcommand{\ec}{\end{center}}
\newcommand{\bqb}{\begin{eqnarray*}}
\newcommand{\eqb}{\end{eqnarray*}}

\begin{document}

\draft
\preprint{May 2004, PM/04-12}

\title{\vspace{1cm}  Special features of heavy
quark-antiquark pair production ratios at LHC
\footnote{Partially supported by EU contract HPRN-CT-2000-00149}}
\author{M. Beccaria$^{a,b}$,
F.M. Renard$^c$ and C. Verzegnassi$^{d, e}$ \\
\vspace{0.4cm}
}

\address{
$^a$Dipartimento di Fisica, Universit\`a di
Lecce \\
Via Arnesano, 73100 Lecce, Italy.\\
\vspace{0.2cm}
$^b$INFN, Sezione di Lecce\\
\vspace{0.2cm}
$^c$ Physique
Math\'{e}matique et Th\'{e}orique, UMR 5825\\
Universit\'{e} Montpellier
II,  F-34095 Montpellier Cedex 5.\hspace{2.2cm}\\
\vspace{0.2cm}
$^d$
Dipartimento di Fisica Teorica, Universit\`a di Trieste, \\
Strada Costiera
 14, Miramare (Trieste) \\
\vspace{0.2cm}
$^e$ INFN, Sezione di Trieste\\
}

\maketitle

\begin{abstract}
We consider the ratio of the cross section of production
of heavy bottom-antibottom pairs ($\sigma_{b\bar b}$)
to the cross section of production of top-antitop pairs 
($\sigma_{t\bar t}$) at LHC.
We show that in this ratio a major fraction of virtual
contributions (including QCD components) disappears, 
leaving a residual dominant electroweak Yukawa effect 
that might be strongly enhanced in the presence of  supersymmetry. 
A quantitative numerical analysis of the size of the effect is
also performed under the assumption of a relatively light 
supersymmetric scenario.
\end{abstract}

\newpage

A widespread hope exists between physicists that the future measurements
at LHC might have a double role, that incorporates the search and
discovery of supersymmetry, via direct production of new sparticles,                                 
 with the subsequent test of the specific details of the 
candidate model through an analysis of its perturbative higher 
order virtual effects. Given
the fact that the latter ones will be, typically, of the relative
percent size, this analysis will have to take properly into account
both the experimental and the theoretical uncertainties of the 
considered process. 
For what concerns the experimental precision, one expects,
typically, a systematic error on e.g. quark-antiquark pair 
production cross sections of a relative few percent from 
luminosity uncertainties. On the theoretical side, 
the major source of uncertainty is coming from
various higher order QCD effects, whose relevance has been underlined
in several recent papers\cite{QCDho}. Under these conditions, the
effect of a relevant virtual genuine electroweak supersymmetric
contribution to a chosen observable might be, least to say, partially
hidden.\par
The aim of this short (and preliminary) paper is that of showing that
there exists a special observable at LHC whose theoretical 
expression is freed from a major
part of the unwanted theoretical uncertainties, 
particularly those of QCD origin. At
the same time, a partial cancellation of experimental systematic
errors is also reasonably expected, which would make this 
observable to play, indeed, a "special" role in the possible 
identification of virtual supersymmetric effects.\par
Our starting point is the observation that, within the set of LEP1
experiments, there was actually one observable that enjoyed 
those combined properties i.e. the ratio $R_b$ of the partial 
widths of $Z$ to $b\bar b$ pairs, $R_b=\Gamma_b/\Gamma_{\rm had}$. 
In fact, owing to the cancellation of several both theoretical
and experimental uncertainties in the ratio,
it became possible to detect from its accurate measurement 
the famous vertex electroweak effect in
the $b\bar b$ decay, proportional to $m^2_t$, and to make strong
predictions concerning the value of the top mass \cite{LEPSLC}.
Given these premises, it seemed almost natural to us to consider  the
generalization of that observable to the LHC situation, with the extra
possibility that is offered by the available creation of top-antitop
pairs. A first possibility would have appeared in fact to consider 
the two ratios of the bottom-antibottom 
and top-antitop pair production cross sections to that of the six 
quark-antiquark pairs. This would be theoretically feasible, 
but from an experimental point of view the precise separation of the
light quark-antiquark component of the denominator from the 
gluon-gluon (or even quark-gluon) ones
seems to us to be problematic, 
and we decided therefore to postpone for the moment the more
complicated theoretical study of the complete light 
quark-antiquark "and similar" channels .       
Since for heavy quarks, on the contrary, both bottom-antibottom 
and top-antitop pair production cross sections will be, 
in principle, measurable with a possible few percent
experimental accuracy \cite{accLHC}, 
the simplest residual choice seemed 
to us that of considering the
ratio                        

\bq
R_{b,t}(s)=[{d\sigma(PP\to b\bar b+...)\over ds}]/
[{d\sigma(PP\to t\bar t+...)\over ds}]
\eq
\noindent
where for a total c.m. squared energy
$S$, the heavy quark $q\bar q$ squared 
invariant mass ($s$)  distribution is given by
\bqa
{d\sigma(PP\to q\bar q+...)\over ds}&=&
{1\over S}~\int^{\cos\theta_{max}}_{\cos\theta_{min}}
d\cos\theta~[~\sum_{ij}~L_{ij}(\tau, \cos\theta)
{d\sigma_{ij\to  q\bar q}\over d\cos\theta}(s)~]
\eqa
\noindent
where $\tau={s\over S}$, and $(ij)$ represent 
all initial $q'\bar q'$ pairs with 
$q'=u,d,s,c,b$ and the initial $gg$ pairs, with the corresponding
luminosities

\bq
L_{ij}(\tau, \cos\theta)={1\over1+\delta_{ij}}
\int^{\bar y_{max}}_{\bar y_{min}}d\bar y~ 
~[~ i(x) j({\tau\over x})+j(x)i({\tau\over x})~]
\eq
\noindent
$i(x)$ being the distributions of the parton $i$ inside the proton
with a momentum fraction,
$x={\sqrt{s\over S}}~e^{\bar y}$, related to the rapidity
$\bar y$ of the $q\bar q$ system.

The limits of integrations for $\bar y$ can be written

\bqa
&&\bar y_{max}=\max\{0, \min\{Y-{1\over2}ln\chi,~Y+{1\over2}ln\chi,
~-ln(\sqrt{\tau})\}\}\nonumber\\
&&
\bar y_{min}= - \bar y_{max}
\eqa
\noindent
where the maximal rapidity is $Y=2$, the  
quantity $\chi$ is related to the scattering angle
in the $q\bar q$ c.m.
\bq
\chi={1+\cos\theta\over1-\cos\theta} 
\eq
and 
\bq
\cos\theta_{min,max}=\mp\sqrt{1-{4p^2_{T,min}\over s}}
\eq
expressed in terms of
the chosen value for $p_{T,min}$.\par 
Note that we might have used a different distribution, like for
instance with respect to transverse momentum, without changing the
conclusions of the analysis. In fact, we have followed here 
(for sake of continuity) the definitions and
conventions of a previous paper where heavy quark-antiquark 
pair production in a supersymmetric scenario was first 
investigated \cite{qqbb}. \par 
For a preliminary particularly simple illustration, 
we shall now concentrate on a specific kinematical
configuration, in particular one where $\sqrt{s}$ lies in the $1$ TeV
range. Under this assumption, it is known from the available
expressions of parton distribution functions \cite{pdf}, and it was
actually numerically checked in \cite{qqbb}, that the dominant
contribution to all the considered quark pair production cross
sections is coming from the initial gluon-gluon state. Therefore we
shall first consider an extreme (preliminary) situation where the
remaining quark-antiquark initial state diagrams can be completely
ignored, and explore the consequences of this assumption.\par
From a theoretical point of view, and to the extent that 
quark masses can be assumed, in the
kinematics of the process, to be much smaller than the 
considered $\sqrt{s}$, we observe that the only appreciable 
difference in the virtual effects
(apart from the standard QED ones that we consider as perfectly
known quantities) is due to the final Yukawa coupling vertex effects,
proportional to the squared quark masses, more precisely to their
ratio to the squared $W$ mass. Actually, the electroweak gauge 
virtual effects are "almost" equal in
the numerator and in the denominator, as
shown in \cite{qqbb}; for what concerns the strong effects,
in particular those of SM QCD origin or
those related to the gluon distribution function inside the proton, 
they are identical in the two cross sections,
apart from the different running mass effects that must be taken 
into account (as it was done in the
LEP1 $R_b$ case), but are not expected to introduce a 
substantial theoretical uncertainty. 
In the ratios that we are considering,
these Yukawa contributions will consequently be  
"clean" (i.e. freed from theoretical uncertainty) electroweak 
virtual effects.  This fact
would probably be of modest relevance within the framework of a
Standard Model analysis, but would acquire a totally different
importance if the model to be tested were a supersymmetric one. To
provide a more quantitative illustration of this statement, we shall
move to the situation, previously considered in \cite{qqbb}, of the
so-called Minimal Supersymmetric Standard Model (MSSM). We shall
assume, as done in that reference, a previous discovery of
Supersymmetry and a suitable "light" scenario in which all relevant
sparticle masses are smaller than, approximately, $400$ GeV. This
assumption is not really a necessary condition, but it allows us to
use the simplified logarithmic expansion of ref\cite{qqbb}. In
particular, we write now the expressions of the relevant differential
cross sections, only explicitely retaining the electroweak SUSY
effects at one loop and omitting SM and SUSY QCD terms that are
canceling in the ratio. 
This leads us to the following formulae:

\bqa
&&{d\sigma^{1~loop}(gg\to b\bar b)\over
dcos\theta}={d\sigma^{Born}(gg\to b\bar b)\over
dcos\theta}~\{~1+{\alpha\over144\pi s^2_Wc^2_W}
(27-22s^2_W)[2ln{s\over M^2_W}-ln^2{s\over M^2_W}]\nonumber\\
&&-~{\alpha \over8\pi s^2_W}[{m^2_t\over M^2_W}
(1+cot^2\beta)+{3m^2_b\over M^2_W}(1+tan^2\beta)][ln{s\over M^2_W}]
~\}
\label{ggbb}\eqa

\bqa
&&{d\sigma^{1~loop}(gg\to t\bar t)\over
dcos\theta}={d\sigma^{Born}(gg\to t\bar t)\over
dcos\theta}~\{~1+{\alpha\over144\pi s^2_Wc^2_W}
(27-10s^2_W)[2ln{s\over M^2_W}-ln^2{s\over M^2_W}]\nonumber\\
&&-~{\alpha \over8\pi s^2_W}[{3m^2_t\over M^2_W}
(1+cot^2\beta)+{m^2_b\over M^2_W}(1+tan^2\beta)][ln{s\over M^2_W}]
~\}
\label{ggtt}\eqa

In all the previous formulae, in the considered $\sqrt{s}$ regime
where only the contribution from t,u Born channels and different gluon
helicities survive (as shown in ref.\cite{qqbb}), the expression of the
Born quantities is the following:

\bq
{d\sigma^{Born}(gg\to q\bar q)\over
dcos\theta}={\pi\alpha^2_s\over4s}[ {u^2+t^2\over 3ut}-
{3(u^2+t^2)\over 4s^2}]
\eq

Using  the previous approximate gluon-gluon dominated formulae, it is
immediate to derive the following simple expression for the
considered  ratio, 
that we write for illustration purposes,
since one easily understands from this equation the main numerical
features that will be illustrated in the following figures:

\bqa
R_{b,t}(s)&=&1-~{\alpha\over12\pi c^2_W} 
[2ln{s\over M^2_W}-ln^2{s\over M^2_W}]\nonumber\\
&&
-~{\alpha\over8\pi s^2_W} [ln{s\over M^2_W}]
[{2m^2_b\over M^2_W}(1+tan^2\beta)-~{2m^2_t\over M^2_W}(1+cot^2\beta)]
\label{rbt}\eqa

Note that the first correction term appearing in the r.h.s 
of the above equation is a residual effect
of the "electroweak gauge" correction present in 
eq.(\ref{ggbb},\ref{ggtt}) which is not completely canceling in the
$R_{b,t}$ ratio, but is numerically irrelevant (of the few permille
order of magnitude).\par

Without entering a precise numerical analysis that will be shown in
the following part of the paper, we observe at this stage that one can 
qualitatively predict an effect in the ratio of Yukawa origin 
which oscillates, depending on the chosen value of $\tan\beta$,
from, roughly, a positive ten percent for small values (around ten) 
to, roughly, a negative five percent for large values 
(around fifty).\par
\begin{center}
\begin{figure}[htb]
\epsfig{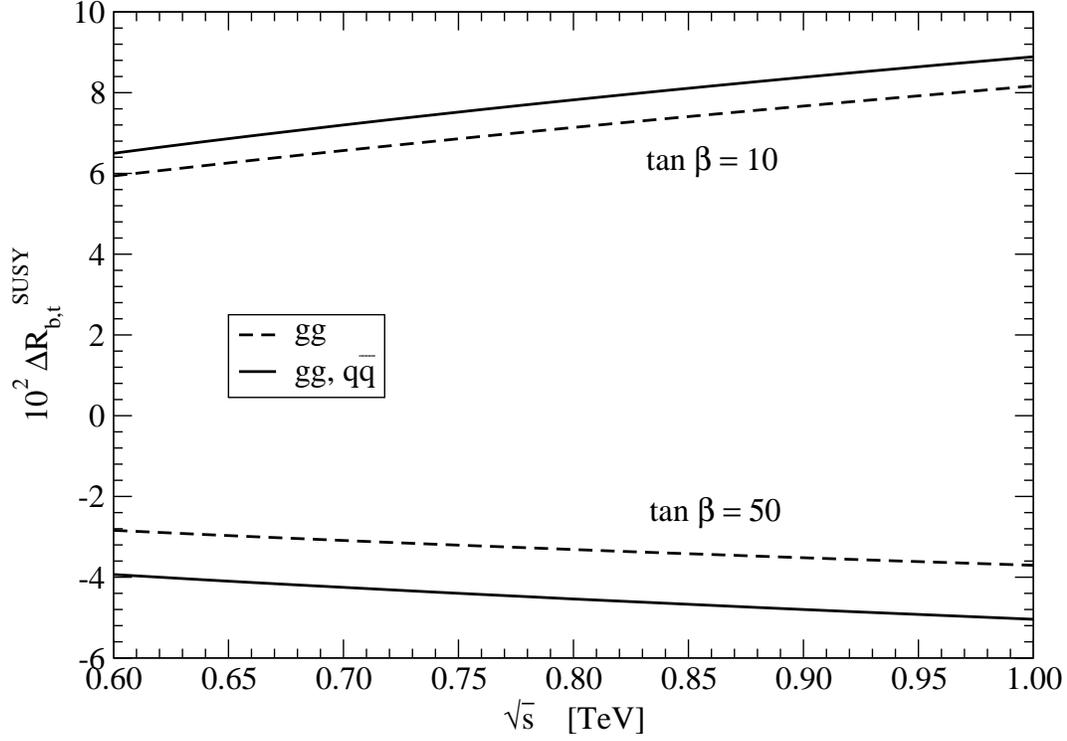}
\vskip 1cm
\caption{Sudakov SUSY effect in $R_{b,t}$: dependence on $\sqrt{s}$. As in the previous
figure, the dashed and solid curves show the effect without and with the contributions coming from 
diagrams with quark-antiquark initial state. $p_{T, \rm min} = 10$ GeV, $\sqrt{S} = 14$ TeV.
 }
\label{fig:energy}
\end{figure}
\end{center}

These  conclusions are, as previously stated, approximate ones
i.e. based on the assumption of gluon-gluon dominance. For what
concerns the considered process, though, we observe that 
all the involved initial quark-antiquark
states SM QCD contributions to the two cross sections are the same, 
with the only exception of those coming from the
t-channel initial bottom-antibottom state that only contributes the 
bottom-antibottom pair production,and we know from an analysis carried
in ref.\cite{qqbb} that this effect is essentially negligible 
already at its Born level. Electroweak
effects of universal kind will not cancel, but, again, one sees from 
ref.\cite{qqbb} that their size remains small (below one percent).  
On the basis of these general facts, we 
expect quite reasonably that the relatively
"large" effects that appear in eq.(\ref{rbt}) 
survive in a more complete
treatment that also includes the quark-antiquark diagrams. In this
paper, we have extended our analysis, retaining the same light
supersymmetric scenario that was considered and taking the extra new
diagrams into account but only within "reasonable" limits. 
To be more precise,
we have now added to the gluon dominated terms those coming from
quark-antiquark states, for whose description we have added to their
Born terms all the MSSM electroweak corrections, 
in particular the extra ones of Yukawa origin,
but not the  QCD terms.\\
The results of our analysis are shown in the next Figures. 
More precisely, we show in Fig.~\ref{fig:tanbeta} the $\tan\beta$ dependence
of the effect at $\sqrt{s}=0.7$ and 1 TeV for $R_{b,t}$  both in 
the gluon-gluon dominance and in the more complete illustrated 
treatment (the lower limit has been chosen following the discussion given in 
ref.~\cite{qqbb}).
In Fig.~\ref{fig:energy} we show the $\sqrt{s}$ dependence of the ratio for two
representative values of $\tan\beta$ and in the energy range 700 GeV - 1 TeV, 
again in the two considered descriptions. 
We have chosen as we said the (realistic) 
value $p_{T,\rm min}=10$ GeV and 
$m_t = 173.8$ GeV, $m_b = 4.25$ GeV.
Different values of $m_b$ would not
change appreciably the small $\tan\beta$ effects, 
but would be more effective in the large $\tan\beta$ region.\par
 As one sees, the effect is practically
identical in the two treatments, and oscillates as we announced 
between positive and negative limits
that might reach the ten percent or five percent peak at the 
extreme  considered $\tan\beta$
values. For a separate experimental accuracy of the two observables 
at the few percent level, this clean genuine electroweak
effect might therefore be seen via a dedicated analysis of 
their ratio.\par
In conclusion, we believe to have shown that the ratio of  
the two heavy quark-antiquark pair production cross sections
$R_{b,t}$ does exhibit, from a theoretical point of
view, "special" features that are, somehow, reminiscent 
of those of the ratio $R_b$ at LEP1. Our analysis has been, 
we stress, preliminary and in particular 
we have adopted a rather convenient
supersymmetric scenario for which a quite simplified 
theoretical logarithmic expansion of so called
"Sudakov kind" at next-to leading order can be adopted. 
But the general property of cancellation of
unwanted theoretical uncertainties will remain true even for 
other scenarios where a less simple description should be
adopted, in particular one for which a complete one-loop expansion 
has to be provided where e.g. other supersymmetric parameters 
different from $\tan\beta$ will enter in a 
more complicated way. A more rigorous work along this direction 
is actually already in progress.  \par

\begin{center}
\begin{figure}[htb]
\epsfig{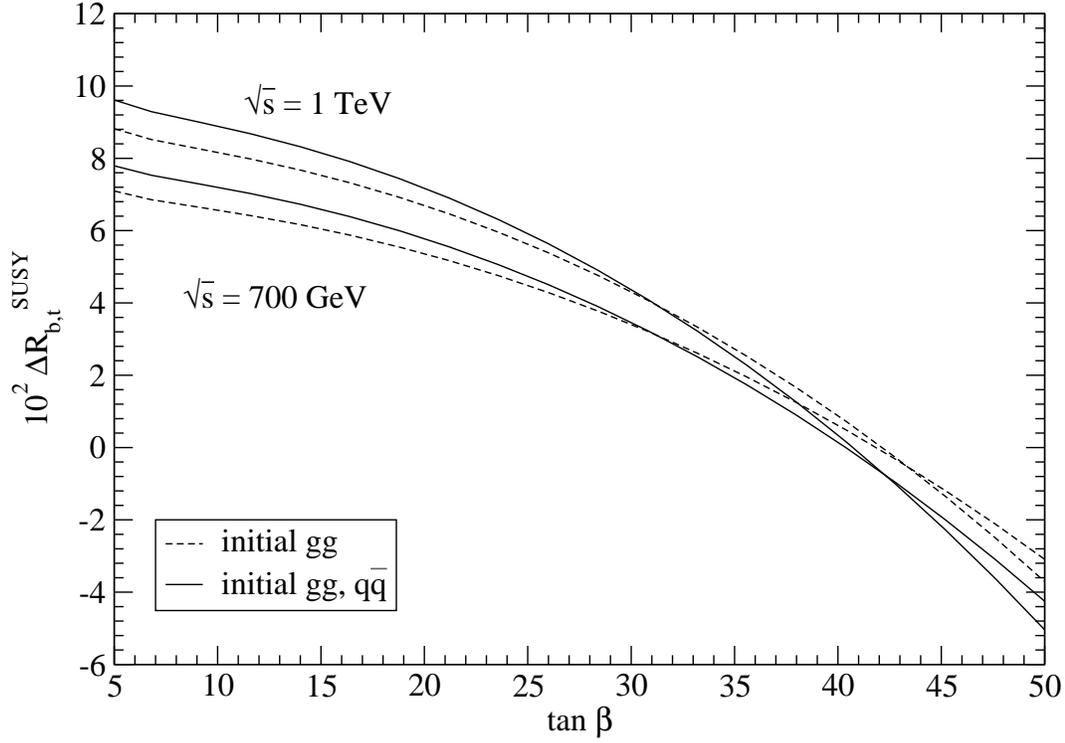}
\vskip 1cm
\caption{Sudakov SUSY effect in $R_{b,t}$: dependence on $\tan\beta$. 
The dashed curve shows the effect in the 
approximate case when we retain the dominating gluon contribution only. The solid
line shows the full effect that includes also the contribution from initial 
quark - antiquark states. $p_{T, \rm min} = 10$ GeV, $\sqrt{S} = 14$ TeV.
 }
\label{fig:tanbeta}
\end{figure}
\end{center}

\end{document}